\documentclass[12pt]{article}

\usepackage[dvips]{graphicx}

\begin{document}
\title{GRB neutrinos, Lorenz Invariance Violation and the influence of background cosmology}
\vspace{0.6 cm}
\author{ Marek Biesiada\\
Department of Astrophysics and Cosmology,\\
University of Silesia,\\
Uniwersytecka 4, 40-007 Katowice, Poland \\
biesiada@us.edu.pl \\[2ex]
 Aleksandra Pi{\'o}rkowska\\
Department of Particle Physics and Field Theory,\\
University of Silesia,\\
Uniwersytecka 4, 40-007 Katowice, Poland \\
apiorko@us.edu.pl}

\date{}

\maketitle 

\begin{abstract}
\noindent

Modern ideas in quantum gravity predict the possibility of Lorenz
Invariance Violation (LIV) manifested e.g. by energy dependent
modification of standard relativistic dispersion relation. In a
recent paper Jacob and Piran proposed that time of flight delays
in high energy neutrinos emitted by gamma ray bursts (GRBs)
located at cosmological distances can become a valuable tool for
setting limits on LIV theories. However, current advances in
observational cosmology suggest that our Universe is dominated by
dark energy with relatively little guidance on its nature thus
leading to several cosmological scenarios compatible with
observations.

In this paper we raise the issue of how important, in the context
of testing LIV theories, is our knowledge of background
cosmological model. Specifically we calculate expected time lags
for high-energy (100 TeV) neutrinos in different cosmological
models. Out of many particular models of dark energy we focus on
five: $\Lambda$CDM, quintessence, quintessence with time varying
equation of state, brane-world and generalized Chaplygin gas model
as representative for various competing approaches.

The result is that uncertainty introduced by our ignorance
concerning the right phenomenological model describing dark energy
dominated universe is considerable and may obscure bounds derived
from studying time delays from cosmological sources.

\end{abstract}


\newpage

\section{Introduction}

Modern approaches to quantum gravity predict the Lorenz Invariance
Violation (LIV thereafter) manifesting itself in particular as an
energy dependent modification of relativistic dispersion relation.
Essentially, additional terms in dispersion relation follow a
power law expansion with respect to $E/E_{QG}$ where $E$ denotes
the particle's (photon's) energy and $E_{QG}$ is the quantum
gravity energy scale. The first guess concerning $E_{QG}$ would be
to assume it being of order of the Planck energy $E_{Pl} = 1.2
\;10^{19}\;GeV$. Although there are suggestions that in some
concrete models (e.g. with large extra dimensions) $E_{QG}$ could
be considerably lower than Planck energy, it is clear that
departures from standard relativistic dispersion relation can be
seen only in high-energy particles or photons.

Several years ago it has been proposed to use astrophysical
objects to look for energy dependent time of arrival delays
\cite{Amelino-Camelia}. Specifically gamma ray bursts (GRBs) being
energetic events visible from cosmological distances are the most
promising sources of constraining LIV theories. Indeed they were
already discussed in this context (quite recently in
\cite{Rodrigues} ) and even used to obtain some constraints
\cite{Ellis}. One is facing here a problem that in the energy
range typical for gamma ray photons the LIV effects are very
subtle. On the other hand, one could imagine looking for TeV
photons which would be produced by GRBs in synchrotron self
Compton mechanism \cite{Tsvi review, Meszaros}. However, the
Universe filled with 2.7 K cosmic microwave background radiation
becomes opaque, via pair production process, to photons with
energies above 10 TeV. This is analogous to GZK threshold for
particles. Despite the fact that LIV theories are often invoked to
resolve the GZK paradox \cite{GZK} and that 20 TeV photons were
reported come from Mk 501 BL Lac object \cite{Mk 501} the use of
very high energy photons from GRBs can be tricky.

Hopefully in a recent paper Jacob and Piran proposed to use high
energy neutrinos instead of photons \cite{Piran Jacob}. Emission
of $100 - 10^4\; TeV$ neutrinos is typically predicted in current
models of GRBs \cite{Tsvi review} and as noticed in \cite{Piran
Jacob} the forthcoming neutrino detectors like Ice Cube are
extremely quiet in this energy range. Measurements of time delay
between prompt gamma ray photons and neutrino signal would open a
new window on exploring LIV theories. Therefore this idea is worth
further consideration. In this paper we discuss the sensitivity of
this setting to the details of cosmological model.

The discovery of accelerated expansion of the Universe
\cite{Perlmutter} introduced the problem of ``dark energy'' in the
Universe which is now one of the most important issues in modern
cosmology. A lot of specific scenarios have been put forward as an
explanation of this puzzling phenomenon. They fall into two broad
categories: searching an explanation among hypothetical candidates
for dark energy (cosmological constant $\Lambda$
\cite{Perlmutter}, quintessence - evolving scalar fields
\cite{Ratra}, Chaplygin gas \cite{Kam}) or modification of gravity
theory (e.g. brane world scenarios \cite{DGP}). We will examine
the LIV induced time delays between prompt photon and neutrino
arrivals in the above mentioned five classes of cosmological
models.

In the next Section we briefly recall phenomenology of distorted
dispersion relation in LIV theories and its consequences for
energy dependent time of arrival delays. Section 2 also briefly
outlines cosmological models considered. Note that brane-world
scenario is not just a candidate for background cosmology, but
also ``represents'' the class of theories in which LIV occurs
\cite{LIV brane}. The final section contains results and
conclusions.

\section{LIV induced time delays in different cosmological models}

Following \cite{Piran Jacob} and for better comparison of results
we assume the modified dispersion relation for neutrinos from GRB
sources in the form:
\begin{equation}
E_{\nu}^2 - p_{\nu}^2 c^2 - m_{\nu}^2 c^4 = \epsilon E_{\nu}^2 \left( \frac{E_{\nu}}{{\xi}_n
E_{QG}}\right)^n \label{dispersion}
\end{equation}
where:$\epsilon = \pm 1$ with $+1$ corresponding to superluminal
and $-1$ to infraluminal motion, ${\xi}_n$ is a dimensionless
parameter. In order to get the results comparable with \cite{Piran
Jacob} we assume $E_{QG}$ equal to the Planck energy, ${\xi}_1 =1$
and ${\xi}_2=10^{-7}$. The dispersion relation (\ref{dispersion})
essentially corresponds to the power-law expansion (see
\cite{Ellis}) so for the practical purposes (due to smallness of
expansion parameter $E/E_{QG}$) only the lowest terms of the
expansion are relevant. Because in some LIV theories the odd power
terms might be forbidden \cite{Burgess} we retain the cases of
$n=1$ and $n=2$.

The relation (\ref{dispersion}) leads to a hamiltonian of the
following form
\begin{equation}
H = \sqrt{\left( p_{\nu}^2 c^2 + m_{\nu}^2 c^4 \right) \lbrack 1 +
\epsilon \left( \frac{E_{\nu}}{{\xi}_n E_{QG}} \right) ^n \rbrack
} \label{hamiltonian}
\end{equation}
Because of the expansion of the Universe, neutrino momentum
$p_{\nu}=p_{\nu} (t)$ is related to the cosmic scale factor $a(t)$
through
\begin{equation}
p_{\nu} (t) = \frac{ p_{\nu} (t_{0}) }{a(t)} \label{relation1}
\end{equation}
Later on the scale factor will be re-expressed in terms of
redshift $z$ which is observable quantity. Similar relation holds
of course for neutrino energy $E_{\nu} = E_{\nu} (t)$.\newline The
time dependent velocity is given by
\begin{equation}
v(t)=\frac{\partial H}{\partial p} \label{velocity}
\end{equation}
From (\ref{velocity}), using hamiltonian (\ref{hamiltonian}),
dispersion relation (\ref{dispersion}) and scale factor dependence
(\ref{relation1})
one can easily obtain (to the lowest order in terms of the observed neutrino
energy $ E_{\nu} (t_{0}) \equiv E_{\nu 0} $ ) that
\begin{equation}
v_{\nu} (t) \simeq \frac{c}{a(t)} \lbrack 1 - \frac{1}{2}
\frac{m_{\nu}^2   c^4}{E_{\nu 0}^2} a^2 (t) + \frac{1}{2} (n+1)
\epsilon \left( \frac{E_{\nu0}}{{\xi}_n E_{QG}} \right)^n
\frac{1}{a^n (t)} \rbrack \label{neutrino velocity1}
\end{equation}

The comoving distance travelled by neutrino from a GRB to the
Earth is defined as
\begin{equation}
r(t) = \int_{t_{emission}}^{t_{0}} v(t) dt \label{distance1}
\end{equation}
Taking into account that $a(t) = \frac{1}{1+z}$ we can express the
above relation (\ref{distance1}) in terms of
 redshift $z$
\begin{equation}
r(z) = \int_{0}^{z} v(z) \frac{dz}{H(z) (1+z)} \label{distance2}
\end{equation}
where for neutrinos we have
\begin{equation}
v(z) \simeq c(1+z) \lbrack 1 - \frac{1}{2} \frac{m_{\nu}^2 c^4}{E_{\nu 0}^2} \frac{1}{(1+z)^2} + \frac{1}{2} (n+1) \epsilon \left( \frac{E_{\nu 0}}{\xi E_{Pl}} \right)^n (1+z)^n \rbrack  \label{neutrino velocity2}
\end{equation}
and $H(z)$, as usually denotes the expansion rate.
Time of flight for neutrinos (i.e. the comoving distance measured
in light years) from GRB source to the Earth is then
\begin{equation}
t_{\nu} 
= \int_{0}^{z} \lbrack 1 - \frac{m_{\nu}^2 c^4}{2 E_{\nu 0}}
\frac{1}{(1+z)^2} + \epsilon \frac{n+1}{2} \left( \frac{E_{\nu
0}}{{\xi}_n E_{QG}} \right)^n (1+z)^n \rbrack \frac{dz}{H(z)}
\label{neutrino time}
\end{equation}
In the first term one easily recognizes the well known time of
flight for prompt (lower energetic) photons
so the time delay due to both, neutrino masses and LIV effects, between a high energy neutrino 
and a low energy prompt photon is equal to
\begin{equation}
\Delta t = 
=  \int_0^z \lbrack  \frac{m_{\nu}^2 c^4}{2 E_{\nu 0}}
\frac{1}{(1+z)^2} - \epsilon \frac{n+1}{2} \left( \frac{E_{\nu
0}}{{\xi}_n E_{QG}}
 \right)^n (1+z)^n \rbrack \frac{dz}{H(z)} \label{time delay}
\end{equation}

In the calculations above we retained the neutrino mass --- it is
massive after all. For the purpose of further calculations we
assume $m_{\nu} = 1\;eV$. However it is evident already from the
formula (\ref{time delay}) that the effect of non-zero mass of the
neutrino is for our purpose negligible --- in perfect accordance
with formulas in \cite{Piran Jacob}.

 The paragraphs
below briefly introduce five types of cosmological models in which
LIV induced time delays will be calculated. We will restrict our
attention to flat models $k=0$ because the flat FRW geometry is
strongly supported by cosmic microwave background radiation (CMBR)
data \cite{Boomerang}.

Friedman - Robertson - Walker model with non-vanishing
cosmological constant and pressure-less matter including the dark
part of it responsible for flat rotation curves of galaxies (the
co called $\Lambda$CDM model) is a standard reference point in
modern cosmology. Sometimes it is referred to as a concordance
model since it fits rather well to independent data (such like
CMBR data, LSS considerations, supernovae data). The cosmological
constant suffers from the fine tuning problem (being constant, why
does it start dominating at the present epoch?) and from the
enormous discrepancy between facts and expectations (assuming that
$\Lambda$ represents quantum-mechanical energy of the vacuum it
should be 55 orders of magnitude larger than observed
\cite{Weinberg}).

Hence another popular explanation of the accelerating Universe is
to assume the existence of a negative pressure component called
dark energy. One can heuristically assume that this component is
described by hydrodynamical energy-momentum tensor with
(effective) cosmic equation of state: $p = w \rho$ where $-1 < w <
-1/3$ \cite{Chiba98}. In such case this component is called
"quintessence".
Confrontation with supernovae and CMBR data
\cite{BeanMelchiorri} led to the constraint $w \leq -0.8$. This
was further improved by combined analysis of SNIa and large scale
structure considerations (see e.g. \cite{Melchiorri}) and from
WMAP data on CMBR \cite{Spergel}. The most recent one comes from
the ongoing ESSENCE supernova survey \cite{Wood-Vasey} and pins
down the equation of state parameter $w$ to the range $-1.07 \pm
0.09 (1 \sigma) \pm 0.12 (systematics)$. For the illustrative
purposes we chose $w = -0.87$ as representing a quintessence model
which is different from cosmological constant and still admissible
by the data.

If we think that the quintessence has its origins in the evolving
scalar field, it would be natural to expect that $w$ coefficient
should vary in time, i.e. $w = w(z).$ An arbitrary function $w(z)$
can be Taylor expanded. Then, bearing in mind that both SNIa
surveys or strong gravitational lensing systems are able to probe
the range of small and moderate redshifts it is sufficient to
explore first the linear order of this expansion. Such
possibility, i.e. $w(z) = w_0 + w_1 z$ has been considered in the
literature (e.g. \cite{Weller}).
Fits to supernovae data performed in the literature suggest
$w_0=-1.5$ and $w_1=2.1$ \cite{JainAlcanizDev} (which is
consistent with fits given in \cite{My JCAP}). Therefore we
adapted these values as representative for this parametrization of
the equation of state.

In the class of generalized Chaplygin gas models matter content of
the Universe consists of pressure-less gas
 with energy density $\rho_m$
representing baryonic plus cold dark matter (CDM) and of the
generalized Chaplygin
 gas with the equation  of state
$p_{Ch} =- \frac{A}{{\rho_{Ch}}^{\alpha}}$ with $0\le \alpha \le
1$, representing dark energy responsible for acceleration of the
Universe. Using the angular size statistics for extragalactic
sources combined with SNIa data it was found in \cite{AlcanizLima}
that in the the $\Omega_m =0.3$ and $\Omega_{Ch}=0.7$ scenario
best fitted values of model parameters are $A_0=0.83$ and
$\alpha=1.$ respectively. Generalized Chaplygin gas models have
been intensively studied in the literature \cite{Makler} and in
particular they have been tested against supernovae data (e.g.
\cite{BiesiadaGodlowski2005} and references therein). Conclusions
from these fits are in agreement with the above mentioned values
of parameters so we used them as representative of Chaplygin Gas
models.

Brane-world scenarios assume that our four-dimensional spacetime
is embedded into 5-dimensional space and gravity in 5-dimensions
is governed by the usual 5-dimensional Einstein-Hilbert action.
The bulk metric induces a 4-dimensional metric on the brane. The
brane induced gravity models \cite{DGP} have a 4-dimensional
Einstein-Hilbert action on the brane calculated with induced
metric. According to this picture, our 4-dimensional Universe is a
surface (a brane) embedded into a higher dimensional bulk
space-time in which gravity propagates. As a consequence there
exists a certain cross-over scale $r_c$ above which an observer
will detect higher dimensional effects. Cosmological models in
brane-world scenarios have been widely discussed in the literature
\cite{Jain}. It has been shown in \cite{Jain} that flat
brane-world Universe with $\Omega_m=0.3$ and $r_c = 1.4
\;H_0^{-1}$ is consistent with current SNIa and CMBR data. Note
that in flat (i.e. $k=0.$) brane-world Universe the following
relation is valid: $\Omega_{r_c} = \frac{1}{4}(1-\Omega_m)^2$.
Futher research performed in \cite{Malcolm2005} based on SNLS
combined with SDSS disfavored flat brane-world models. More recent
analysis by the same authors \cite{Malcolm2007} using also ESSENCE
supernovae sample and CMB acoustic peaks lead to the conclusion
that flat brane-world scenario is only slightly disfavored,
although inclusion of baryon acoustic oscillation peak would ruled
it out. Despite this interesting debate we use flat brane-world
scenario with $\Omega_m=0.3$ for illustration.

Expansion rates $H(z)  = {\dot a}/{a}$ (equivalent to Friedman
equation) for the models studied are shown in Table 1.

\section{Results and conclusions}

We have calculated time delays of $100 \;TeV$ neutrinos as a
function of redshift (see equation (\ref{time delay})) in
different dark energy scenarios described above and for LIV
theories with $n=1$, $\xi_1=1$ and $n=2$, $\xi_2 = 10^{-7}$
respectively. They are summarized in Figure 1. Redshift range from
$z=0$ to $z=6$ represents the depth of GRB surveys \cite{Tsvi
review} and hence reflects the range of distances from which one
might expect the high energy neutrinos to come. For better
resolution we displayed the same information in Figure 2 but in a
restricted range of redshifts. GRB sample with measured redshifts
has a mode at about $z \sim 1.5$. Therefore a range from $z=2$ to
$z=3$ in some sense also represents the most likely distance from
a potential source of high energy neutrinos. Figure 3, at last,
displays the energy dependence of time of flight delay for the
source located at $z=3$.

One can see noticeable differences between time delays calculated
for different background cosmologies. $\Lambda$CDM model and
quintessence model (with $w$ parameter best fitted to current SNIa
and CMBR data) introduce negligible confusion to time delays.
Brane world models (i.e. the class representative of theories in
which LIV is expected) and Chaplygin gas scenario predict time
delays considerably lower than in $\Lambda$CDM cosmology. For
example the differences in time delays of $100 \; TeV$ neutrino
from a source at $z=3$ between $\Lambda$CDM and Chaplygin gas
model is almost 3 hours for $n=2$ LIV theories and 43 minutes for
$n=1$ LIV theories. Respective differences between $\Lambda$CDM
and brane world models is almost 1 hour for $n=2$ and 16 minutes
for $n=1$. These systematic differences get higher with redshift.
The most pronounced is the difference in time delays between
$\Lambda$CDM and Var Quintessence (i.e. the model with linear
$w(z)$ functions with parameters best fitted to SNIa). The
resulting mismatch between predicted time delays (from a source at
$z=3$) ranges from 1.25 hour in $n=1$ theories to 6 hours in $n=2$
LIV theories. Respective values for more distant source (at $z=6$)
are almost 4 hours ($n=1$) and 27.5 hours.

Our results indicate that our ignorance concerning true model of
dark energy in the universe is not able to spoil the utility of
time delays in discriminating between $n=1$ and $n=2$ classes of
LIV theories. However in each class of LIV theories it introduces
an uncertainty at the level from 7\% ($\Lambda$CDM, Quintessence,
Chaplygin, braneworld) up to 35\% (Var Quintessence)  for sources
at $z=3$ (i.e. the most likely located GRBs). This translates into
ranges 7\% -- 35\% and 14\% -- 70\% uncertainty for inferred
bounds on $\xi_n E_{QG}$ in $n=1$ and $n=2$ cases respectively.
For more distant sources this is respectively higher.

Therefore the conclusion is that better understanding of dark
energy dominated Universe is crucial for testing LIV theories with
cosmological sources like GRBs. Theoretically one may also invert
this argument by saying that if LIV dispersion relation was proven
experimentally and its parameters were constrained then time
delays from GRBs could become a new kind of cosmological test.

\newpage

\begin{table}[hbp]

\caption{Expansion rates $H(z)$ in four models tested. The
quantities $\Omega_i$ represent fractions of critical density
currently contained in energy densities of respective components
(like clumped pressure-less matter, $\Lambda$, quintessence,
Chaplygin gas or brane effects).}
\bigskip

 \begin{tabular}{|c|c|} \hline
 Model & Cosmological expansion rate $H(z)$ (the Hubble function). \\
\hline
  $\Lambda$CDM & $H^2(z) = H^2_0 \left[ \Omega_m \; (1+z)^3 + \Omega_{\Lambda} \right]$  \\
Quintessence & $H^2(z) = H^2_0 \left[ \Omega_m \; (1+z)^3 +
\Omega_Q \; (1+z)^{3(1+w)} \right]
$ \\
Var Quintessence & $H^2(z) = H^2_0 \left[ \Omega_m \; (1+z)^3 +
\Omega_Q \;
(1+z)^{3(1+w_0-w_1)}\;\exp(3 w_1 z) \right]$  \\
Chaplygin Gas & $H(z)^2 = H_0^2 \left[ \Omega_{m} (1+z)^3 +
\Omega_{Ch} \left(A_0 + (1 - A_0)(1+z)^{3(1+ \alpha)}
\right)^{{1\over 1+\alpha}} \right]
$ \\
Braneworld & $H(z)^2 = H_0^2 \left[ (\sqrt{ \Omega_{m} (1+z)^3 +
\Omega_{r_c} }
+ \sqrt{\Omega_{r_c}} )^2 \right]$   \\
\hline
\end{tabular}

\end{table}

\begin{figure}[hbp]
\begin{center}
\includegraphics[scale=1.3]{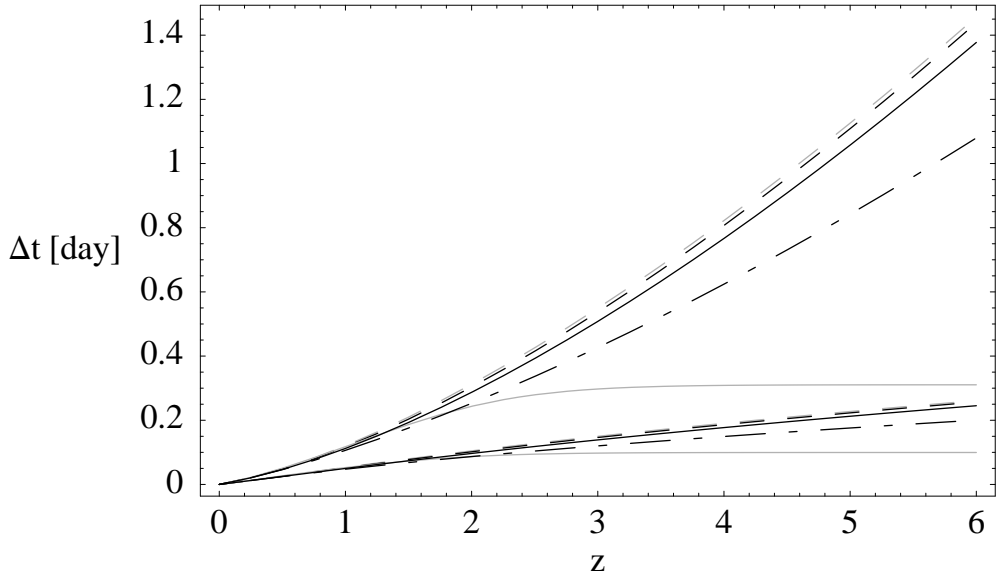}
\caption{
 Observed time delays for $100\; TeV$ neutrinos as a function of
 redshift in different dark energy scenarios ($\Lambda$CDM ---light gray dashed line, quintessence --- black dashed line, quintessence with varying E.O.S.
 --- light gray solid line, brane world model --- black solid line and Chaplygin gas
 scenario --- dot-dashed line). Upper curves correspond to $n=2$, $\xi_2
 = 10^{-7}$, lower curves correspond to $n=1$, $\xi_1=1$.
}
\end{center}
\end{figure}

\begin{figure}[hbp]
\begin{center}
\includegraphics[scale=1.3]{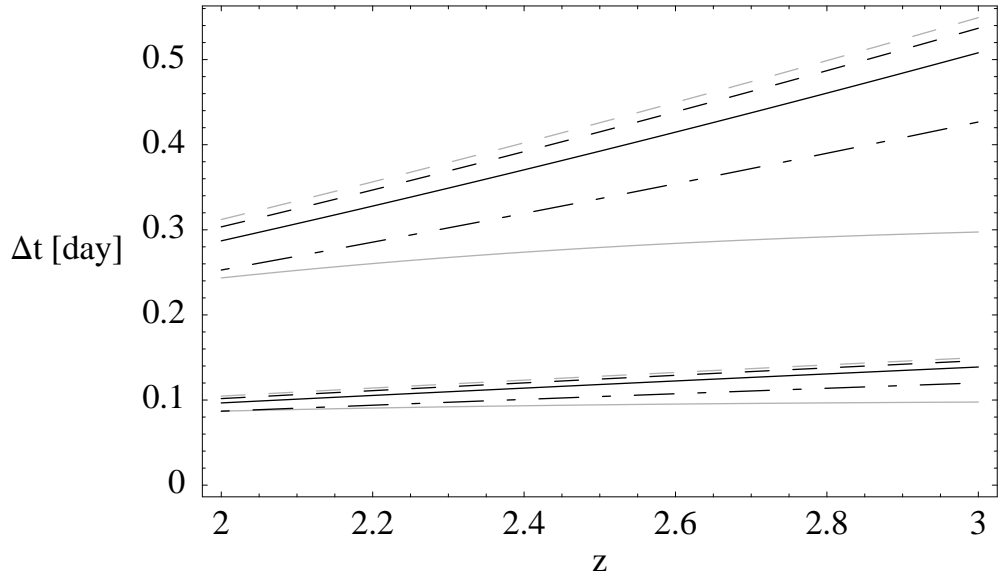}
\caption{
The same as Fig.1, in a restricted redshift
range corresponding to the mode of GRB distribution.
}
\end{center}
\end{figure}


\begin{figure}[hbp]
\centering
\includegraphics[scale=1.3]{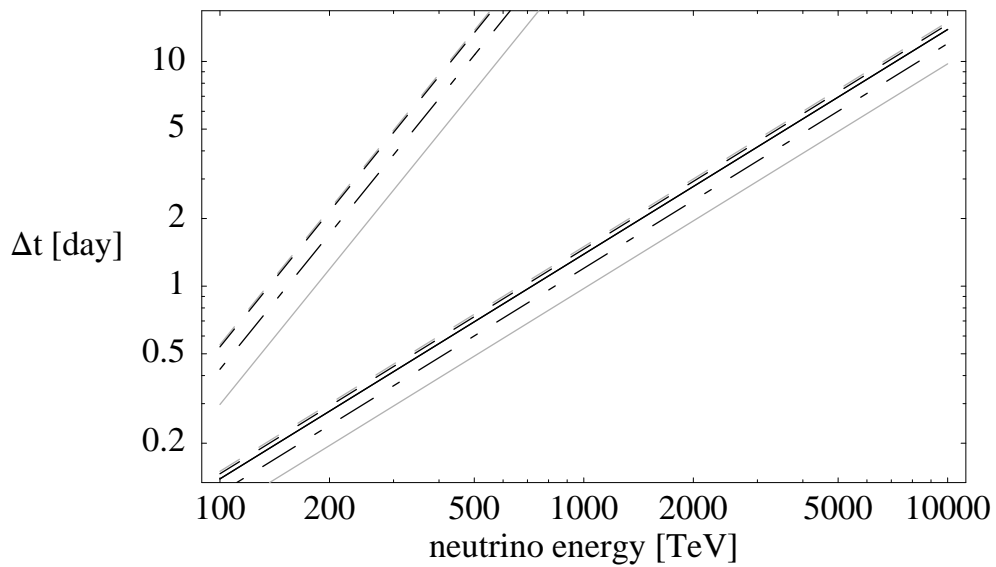}
\caption{Time delays a a function of neutrino energy in different dark
energy scenarios for a source located at $z=3$. Left (steeper)
family of curves corresponds to $n=2$, $\xi_2=10^{-7}$ LIV
theories, right family corresponds to $n=1$, $\xi_1=1$ LIV
theories.}
 \end{figure}



\newpage
\begin{thebibliography}{X}

\bibitem{Amelino-Camelia}
G. Amelino-Camelia, J.R. Ellis, N.E. Mavromatos, D.V. Nanopoulos
and S. Sarkar, Nature {\bf 393}, 763, 1998

\bibitem{Rodrigues}
M. Rodriguez Martinez and T. Piran, 
J Cosmol. Astropart. Phys. 0604 (2006) 006 astro-ph/0601219

\bibitem{Ellis}
J.R. Ellis, N.E. Mavromatos, D.V. Nanopoulos and A.S. Sakharov,
Astron. Astrophys. {\bf 402}, 409, 2003 \\
S.E. Boggs, C.B. Wunderer, K. Hurley and W. Coburn, ApJ {\bf 611},
L77--L80, 2004

\bibitem{Tsvi review}
T. Piran, Rev. Mod. Phys. {\bf 76}, 1143, 2004

\bibitem{Meszaros}
P. Meszaros and M.J. Rees, M.N.R.A.S. {\bf 269}, L41, 1994 \\
P. Meszaros, M.J. Rees and H. Papathanassiou, ApJ {\bf 432}, 181,
1994

\bibitem{GZK}
S. Coleman and S.L. Glashow, Phys. Rev. D {\bf 59}, 116008, 1999
\\
O. Bertolami and C.S. Carvalho, Phys. Rev. D {\bf 61}, 103002,
2000

\bibitem{Mk 501}
G. Amelino-Camelia and T. Piran, Phys. Rev. D {\bf 64}, 036005,
2001 \\
T. Kifune, ApJL {\bf 518}, L21, 1999

\bibitem{Piran Jacob}
U. Jacob and T. Piran, ``GRBs neutrinos as a tool to explore
quantum gravity induced Lorenz violation''hep-ph/0607145 v1

\bibitem{Perlmutter}
S. Perlmutter, G. Aldering, G. Goldhaber, et al., Astrophys. J  {\bf 517}, 565, 1999 \\
A. Riess, A.V. Filipenko and P. Challis, et al., Astron. J {\bf
116}, 1009, 1998

\bibitem{Ratra}
B. Ratra and P.J.E. Peebles, Phys.Rev.D {\bf 37}, 3406, 1988\\
R.R. Caldwell, R. Dave and P.J. Steinhardt, Phys. Rev. Lett. {\bf 75}, 2077, 1995\\
J. Frieman, C. Hill, A. Stebbins and I. Waga, Phys. Rev. Lett.
{\bf
75}, 2077, 1995 \\
R. Caldwell,R. Dave, and P.J. Steinhardt, Phys. Rev. Lett. {\bf 80},1582, 1998\\
I. Zlatev, L. Wang and P. J. Steinhardt, Phys. Rev.Lett. {\bf 82},
896, 1999

\bibitem{Kam}
A. Kamenshchik, V. Moschella and V. Pasquier, Phys.Lett. B, {\bf 511}, 256, 2000; \\
J.C. Fabris, S.V.B. Gon{\c c}alves and P.E. de Souza,
Gen.Rel.Grav. {\bf 34}, 53, 2002

\bibitem{DGP}
G.Dvali, G.Gabadadze and M.Porrati, Phys. Lett.B {\bf 485}, 208, 2000\\
G.Dvali and G.Gabadadze, Phys.Rev. {\bf D63}, 065007, 2001

\bibitem{LIV brane}
C. Csaki, J. Erlich and C. Grojean,
Gen.Rel.Grav. {\bf 33},1921, 2001\\
 O. Bertolami and C. Carvalho,
Phys. Rev. D {\bf 74}, 084020, 2006

\bibitem{Burgess}
C.P. Burgess, J.M. Cline, E. Filotas, J. Matias and G.D. Moore,
JHEP, 0203, 043, 2002

\bibitem{Riess 2004}
A.G. Riess et al. [Supernova Search Team Collaboration],Astrophys.
J. {\bf 607}, 665, 2004

\bibitem{Boomerang}
A. Benoit, et al., Astron.Astrophys. {\bf 399}, L25-L30, 2003

\bibitem{Weinberg}
S. Weinberg, Rev. Mod. Phys. {\bf 61}, 1, 1989

\bibitem{Chiba98}
T. Chiba, N. Sugiyama and T. Nakamura, M.N.R.A.S. {\bf 301}, 72,
1998\\
M.S. Turner and M. White, Phys.Rev.D {\bf 56}, 4439, 1997

\bibitem{BeanMelchiorri}
R. Bean and A. Melchiorri, Phys.Rev. D {\bf 65}, 041302, 2002

\bibitem{Melchiorri}
A. Melchiorri, L. Mersini, C.J. ${\rm{\ddot O}}$dman and M.
Trodden, Phys.Rev. D, {\bf 68},43509, 2003

\bibitem{Spergel}
D. Spergel et al. Astrophys. J Suppl. {\bf 148}, 175, 2003

\bibitem{Wood-Vasey}
W.M. Wood-Vasey et al. (2007) astro-ph/0701041

\bibitem{Weller}
J. Weller and A. Albrecht, Phys.Rev.Lett. {\bf 86}, 1939-1942, 2001\\
I. Maor, R. Brustein and P.J. Steinhardt, Phys.Rev.Lett. {\bf 86},
6-9, 2001

\bibitem{JainAlcanizDev}
D. Jain, J.S. Alcaniz and A. Dev, Nucl. Phys. B {\bf 732}, 379,
2006

\bibitem{My JCAP}
M. Biesiada, J.Cosmol. Astropart. Phys. 02 (2007) 003


\bibitem{AlcanizLima}
J.S. Alcaniz and J.A.S. Lima, Astrophys. J {\bf 618}, 16, 2005

\bibitem{Makler}
M. Makler, S.Q. de Oliveira and I. Waga, Phys.Lett. B {\bf 555}, 1, 2003\\
P.P. Avelino,  L.M.G. Be{\c c}a, J.P.M. de Carvalho, C.J.A.P.
Martins and P. Pinto, Phys. Rev. D {\bf 67}, 023511, 2003

\bibitem{BiesiadaGodlowski2005}
M. Biesiada, W. God{\l}owski and M. Szyd{\l}owski, Astrophys. J
{\bf 622}, 28--38, 2005

\bibitem{Jain}
D. Jain, A. Dev, and J.S. Alcaniz, Phys.Rev. {\bf D66}, 083511, 2002\\
J.S. Alcaniz, D. Jain and A. Dev, Phys.Rev. {\bf D66}, 067301, 2002\\
C. Deffayet, S.J. Landau, J. Raux, M. Zaldarriaga and P. Astier,
Phys.Rev. {\bf D66}, 024019, 2002

\bibitem{Malcolm2005}
M. Fairbairn and A. Goobar, Phys.Lett. {\bf B 642}, 432, 2006

\bibitem{Malcolm2007}
S. Rydbeck, M. Fairbairn and A. Goobar, ``Testing the DGP model
with ESSENCE'', 2007 astro-ph/0701495v1


\end{thebibliography}
\end{document}